\documentclass{article}
\usepackage{geometry}
\usepackage[utf8]{inputenc}
\usepackage{cite}
\usepackage{url}
\usepackage{amssymb,amsmath,amsthm}
\usepackage{bm}
\usepackage{graphicx}
\usepackage{enumerate}
\usepackage{microtype}
\usepackage{color}
\usepackage{hyperref}
\usepackage[ruled,linesnumbered,vlined]{algorithm2e}
\usepackage{multirow}
\usepackage{array}
\newcolumntype{C}[1]{>{\centering\let\newline\\\arraybackslash\hspace{0pt}}m{#1}}

\definecolor{webgreen}{rgb}{0,.35,0}
\definecolor{webbrown}{rgb}{.6,0,0}
\definecolor{RoyalBlue}{rgb}{0,0,0.9}
\definecolor{purp}{rgb}{0.6,0.05,0.8}
\definecolor{ora}{rgb}{0.7,0.35,0.02}

\hypersetup{
   colorlinks=true, linktocpage=true, pdfstartview=FitV,
   breaklinks=true, pdfpagemode=UseNone, pageanchor=true, pdfpagemode=UseOutlines,
   plainpages=false, bookmarksnumbered, bookmarksopen=true, bookmarksopenlevel=1,
   hypertexnames=true, pdfhighlight=/O,
   urlcolor=webbrown, linkcolor=RoyalBlue, citecolor=webgreen,
   pdfauthor={Gary P. T. Choi},
   pdfsubject={Efficient conformal parameterization of multiply-connected surfaces using quasi-conformal theory}
}

\newcounter{exctr}

\title{Efficient conformal parameterization of multiply-connected surfaces using quasi-conformal theory}

\author{Gary P. T. Choi\\
\\
\footnotesize{Department of Mathematics, Massachusetts Institute of Technology, Cambridge, MA, USA}\\
\footnotesize{E-mail: ptchoi@mit.edu}
}
\date{}

\begin{document}

\maketitle
\begin{abstract}
Conformal mapping, a classical topic in complex analysis and differential geometry, has become a subject of great interest in the area of surface parameterization in recent decades with various applications in science and engineering. However, most of the existing conformal parameterization algorithms only focus on simply-connected surfaces and cannot be directly applied to surfaces with holes. In this work, we propose two novel algorithms for computing the conformal parameterization of multiply-connected surfaces. We first develop an efficient method for conformally parameterizing an open surface with one hole to an annulus on the plane. Based on this method, we then develop an efficient method for conformally parameterizing an open surface with $k$ holes onto a unit disk with $k$ circular holes. The conformality and bijectivity of the mappings are ensured by quasi-conformal theory. Numerical experiments and applications are presented to demonstrate the effectiveness of the proposed methods.
\end{abstract}

\section{Introduction}
The goal of surface parameterization is to map a surface in $\mathbb{R}^3$ onto a simple standardized domain. Over the past few decades, surface parameterization algorithms have been extensively studied~\cite{floater2005surface,sheffer2006mesh,hormann2007mesh}. In general, any parameterization will unavoidably induce angle and/or area distortions. Therefore, it is common to consider \emph{conformal parameterizations}, which preserve angles and hence the local geometry of the surfaces. Existing conformal parameterization methods include harmonic energy minimization~\cite{pinkall1993computing,gu2004genus} and its linearizations~\cite{choi2015flash,choi2016spherical,choi2018linear}, least-square conformal map (LSCM)~\cite{levy2002least}, discrete natural conformal parameterization (DNCP)~\cite{desbrun2002intrinsic}, holomorphic 1-form~\cite{gu2003global}, Yamabe flow~\cite{luo2004combinatorial}, angle-based flattening (ABF)~\cite{sheffer2001parameterization,sheffer2005abf}, circle patterns~\cite{kharevych2006discrete}, discrete conformal equivalence~\cite{springborn2008conformal}, Ricci flow~\cite{jin2008discrete,yang2009generalized,yang2008optimal,zhang2015survey}, spectral conformal map~\cite{mullen2008spectral}, curvature prescription~\cite{ben2008conformal}, zipper algorithms~\cite{marshall2007convergence,choi2020parallelizable}, boundary first flattening~\cite{sawhney2017boundary}, conformal energy minimization~\cite{yueh2017efficient} etc. In recent years, quasi-conformal theory has emerged as a useful tool for the development of surface parameterization methods~\cite{choi2015fast,meng2016tempo} with applications to image and video processing~\cite{lui2013texture,yung2018efficient}, geometry processing and graphics~\cite{lipman2012bounded,weber2012computing,wong2014computation,lui2014teichmuller,choi2016fast}, metamaterial design~\cite{choi2019programming}, medical visualization~\cite{zeng2010supine,choi2017conformal} and biological shape analysis~\cite{choi2018planar,choi2020tooth,choi2020shape}. However, most of the above-mentioned conformal parameterization methods only work for simply-connected surfaces, which do not contain any holes. 

For multiply-connected surfaces with annulus or poly-annulus topology, the computation of conformal maps is more complicated. Some earlier works have considered mapping a multiply-connected open surface onto a circular domain with concentric circular slits~\cite{yin2008slit,wang2008conformal}. Also, by the Koebe’s uniformization theorem, any multiply-connected open surface with $k$ holes can be conformally mapped to a unit disk with $k$ circular holes~\cite{koebe1910konforme}. Based on this remarkable result, a few parameterization algorithms have been developed for multiply-connected open surfaces using Ricci flow~\cite{jin2008discrete}, holomorphic 1-form~\cite{zeng2009generalized}, Laurent series~\cite{kropf2014conformal}, Beltrami energy minimization~\cite{ho2016qcmc}, discrete conformal equivalence~\cite{bobenko2016discrete} etc.

\begin{figure}[t]
    \centering
    \includegraphics[width=\textwidth]{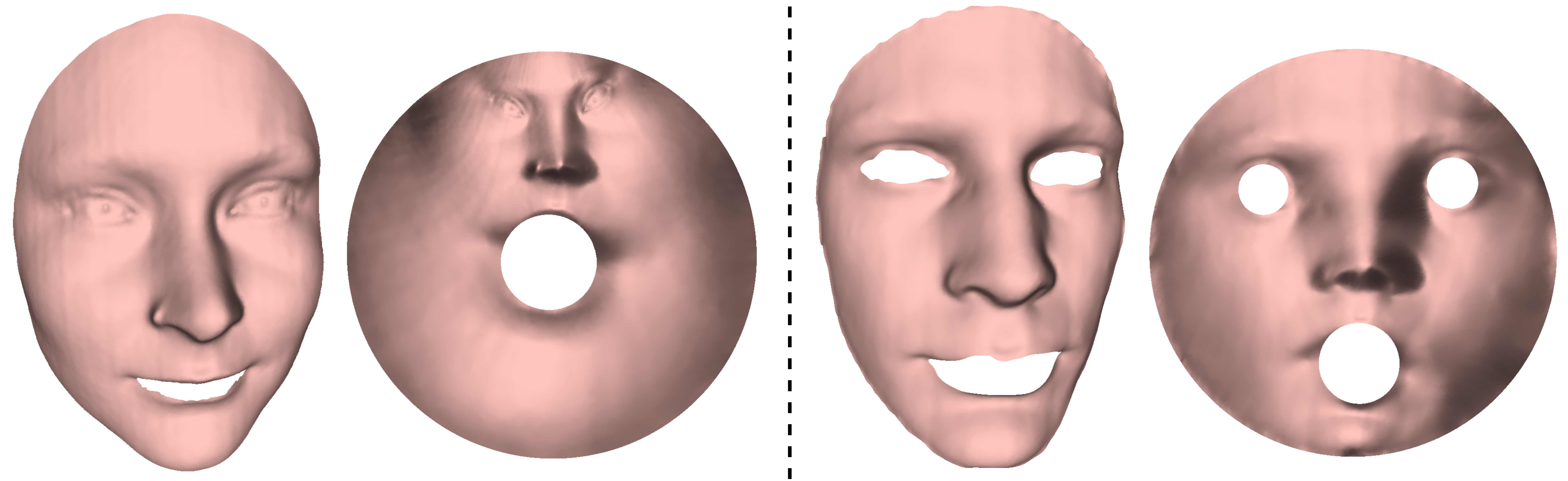}
    
    \caption{Conformal parameterizations of multiply-connected surfaces achieved by our proposed methods. (Left) The conformal parameterization of an open surface with one hole onto an annulus by our Annulus Conformal Map (ACM) algorithm. (Right) The conformal parameterization of a multiply-connected open surface onto a unit disk with circular holes by our Poly-Annulus Conformal Map (PACM) algorithm.}
    \label{fig:illustration}
\end{figure}

In this work, we propose two novel algorithms for computing the conformal parameterization of multiply-connected surfaces using quasi-conformal theory. We first propose an efficient method for conformally mapping an open surface with one hole (i.e. a topological annulus) to an annulus domain with unit outer radius (Fig.~\ref{fig:illustration}, left). We then utilize this method to develop another fast algorithm for conformally mapping a multiply-connected open surface with $k$ holes (i.e. a topological poly-annulus) to a unit disk with $k$ circular holes (Fig.~\ref{fig:illustration}, right). With the aid of quasi-conformal theory, we can effectively achieve the conformality and bijectivity of the parameterizations.

The rest of the paper is organized as follows. In Section~\ref{sect:background}, we review the concepts of conformal and quasi-conformal maps. In Section~\ref{sect:main}, we describe our proposed methods for the conformal parameterization of multiply-connected surfaces. In Section~\ref{sect:results}, we demonstrate the effectiveness of our parameterization methods using numerical experiments. Applications of the proposed methods are explored in Section~\ref{sect:application}. We conclude the paper and discuss possible future directions in Section~\ref{sect:discussion}.

\section{Mathematical background}\label{sect:background}
In this section, we review some mathematical concepts related to our work. Readers are referred to~\cite{lehto1973quasiconformal,gardiner2000quasiconformal,ahlfors2006lectures} for more details.
\subsection{Conformal map}
Let $f: \mathbb{C}\rightarrow\mathbb{C}$ be a map with $f(z) = f(x,y) = u(x,y)+iv(x,y)$, where $u,v$ are real-valued functions. $f$ is said to be a \emph{conformal map} if it satisfies the Cauchy--Riemann equations:
\begin{equation}
\left\{\begin{array}{ll}
    \dfrac{\partial u}{\partial x} &= \dfrac{\partial v}{\partial y},\\
    \dfrac{\partial u}{\partial y} &= -\dfrac{\partial v}{\partial x}.
\end{array} \right.
    \label{eqt:cauchyriemann}
\end{equation}

\emph{M\"obius transformations} are a special class of conformal maps on the complex plane. Mathematically, a M\"obius transformation $f:\mathbb{C}\rightarrow\mathbb{C}$ is in the form
\begin{equation}
    f(z) = \dfrac{az+b}{cz+d}, 
\end{equation}
with $a,b,c,d\in\mathbb{C}$ satisfying $ad-bc\neq 0$.

\subsection{Quasi-conformal map}
\emph{Quasi-conformal maps} are a generalization of conformal maps. Mathematically, a mapping $f:\mathbb{C} \to \mathbb{C}$ is said to be a quasi-conformal map if it satisfies the Beltrami equation
\begin{equation}\label{eqt:beltramieqt}
 \frac{\partial f}{\partial \bar{z}} = \mu_f(z) \frac{\partial f}{\partial z}
\end{equation}
for some complex-valued function $\mu_f$ with $\|\mu_f\|_\infty<1$, where the complex derivatives are given by
\begin{equation}
     \frac{\partial f}{\partial \bar{z}} = f_{\bar{z}} = \frac{1}{2}\left(\frac{\partial f}{\partial x} + i \frac{\partial f}{\partial y}\right)\ \ \text{ and } \ \  \frac{\partial f}{\partial z} = f_{z} = \frac{1}{2}\left(\frac{\partial f}{\partial x} - i\frac{\partial f}{\partial y}\right).
\end{equation} 
Here, $\mu_f$ is called the \emph{Beltrami coefficient} of $f$. Note that if $\mu_f \equiv 0$, then Eq.~\eqref{eqt:beltramieqt} becomes the Cauchy--Riemann equations~\eqref{eqt:cauchyriemann} and hence $f$ is conformal.

\begin{figure}[t]
    \centering
    \includegraphics[width=\textwidth]{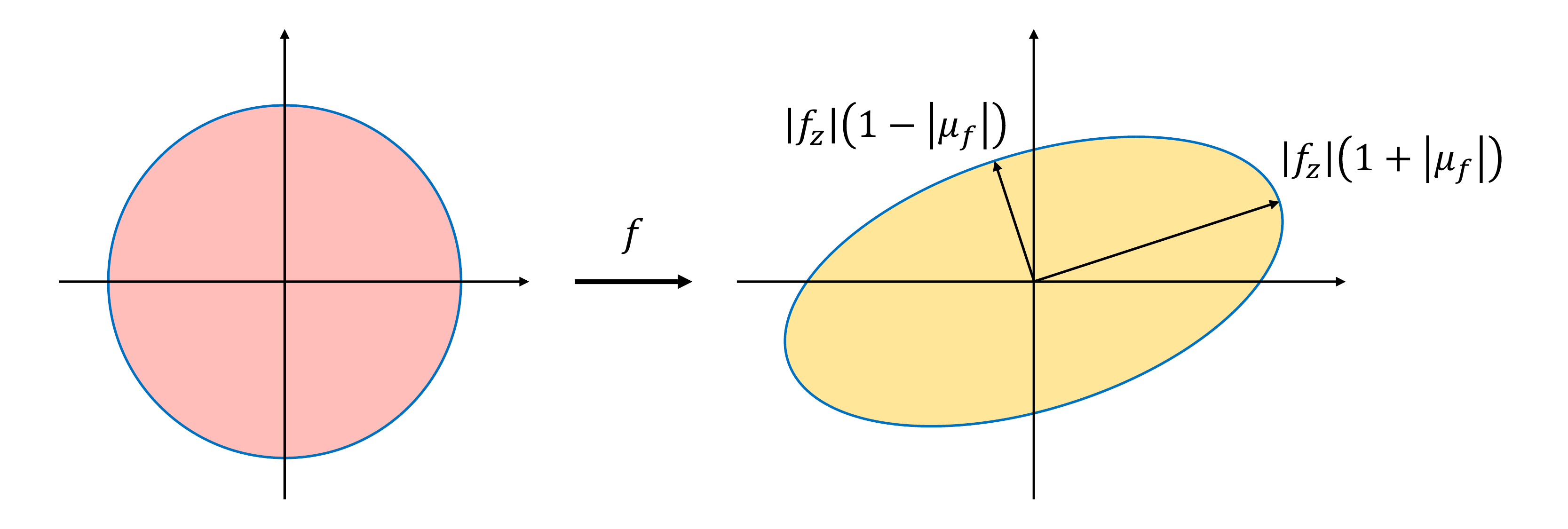}
    \caption{An illustration of quasi-conformal maps. Under a quasi-conformal map $f$, an infinitesimal circle is mapped to an infinitesimal ellipse with the maximum magnification $|f_z|(1+|\mu_f|)$ and the maximum shrinkage $|f_z|(1-|\mu_f|)$.}
    \label{fig:qc_figure}
\end{figure}

Intuitively, conformal mappings map infinitesimal circles to infinitesimal circles, while quasi-conformal mappings map infinitesimal circles to infinitesimal ellipses with bounded eccentricity. To see this, consider the first order approximation of $f$ around a point $z_0 \in \mathbb{C}$:
\begin{equation}
\begin{split}
f(z) &\approx f(z_0) + f_z(z_0) (z-z_0) + f_{\overline{z}}(z_0) \overline{z-z_0} = f(z_0) + f_z(z_0)\left(z-z_0 + \mu_f(z_0) \overline{z-z_0}\right).
\end{split}
\end{equation}
This indicates that an infinitesimal circle centered at $z_0$ is mapped to an infinitesimal ellipse centered at $f(z_0)$, with the maximum magnification $|f_z(z_0)|(1+|\mu_f(z_0)|)$ and the maximum shrinkage $|f_z(z_0)|(1-|\mu_f(z_0)|)$ (see Fig.~\ref{fig:qc_figure} for an illustration). The aspect ratio of the ellipse is then given by $\frac{1+|\mu_f(z_0)|}{1-|\mu_f(z_0)|}$. Therefore, the Beltrami coefficient effectively captures the conformal distortion of its associated mapping.

The Beltrami coefficient is also closely related to the bijectivity of the mapping. Note that if $f(x,y) = u(x,y) + i v(x,y)$, then the Jacobian $J_f$ of $f$ is given by
\begin{equation}
\begin{split}
J_f &= u_x v_y - v_x u_y \\
&= \frac{1}{4}\left( (u_x+v_y)^2 + (v_x - u_y)^2 - (u_x - v_y)^2 - (v_x + u_y)^2\right) \\
&= \frac{1}{4}\left(|(u_x + iv_x) - i (u_y + iv_y)|^2 - |(u_x + iv_x) + i (u_y + iv_y)|^2\right) \\
&= |f_z|^2 - |f_{\bar{z}}|^2 \\
&= |f_z|^2 (1-|\mu_f|^2).
\end{split}
\end{equation}
Therefore, we have the following result:

\noindent \textbf{Theorem 1}. If $f$ is a $C^1$ map satisfying $\|\mu_f\|_{\infty} <1$, then $f$ is bijective.\\


Besides, the Beltrami coefficient of a composition of two quasi-conformal maps can be expressed explicitly. Let $f: \Omega_1 \to \Omega_2$ and $g: \Omega_2 \to \Omega_3$ be two quasi-conformal maps. The Beltrami coefficient of the composition map $g \circ f$ is given by
\begin{equation}
\mu_{g \circ f}  = \frac{\mu_f+(\overline{f_z}/f_z) (\mu_g \circ f)}{1+(\overline{f_z}/f_z)  \overline{\mu_f} (\mu_g \circ f)}.
\end{equation}
In particular, if $\mu_{f^{-1}} \equiv \mu_g$, we have
\begin{equation}
   \mu_g \circ f =  \mu_{f^{-1}} \circ f = -(f_z / \overline{f_z}) \mu_f
\end{equation}
and hence
\begin{equation}\label{eqt:composition}
\mu_{g \circ f} \equiv \frac{\mu_f+(\overline{f_z}/f_z) ((-f_z/\overline{f_z}) \mu_f)}{1+(\overline{f_z}/f_z)  \overline{\mu_f} ((-f_z/\overline{f_z}) \mu_f)} \equiv 0,
\end{equation}
which implies that the composition map $g \circ f$ is conformal. This suggests that one can eliminate the conformal distortion of a quasi-conformal map by composing it with another quasi-conformal map with the same Beltrami coefficient, provided that the boundary constraint is admissible. This idea of quasi-conformal composition~\cite{choi2015flash} will be used in our proposed methods for the computation of conformal parameterizations.

While the above concepts are introduced in terms of mappings on the complex plane, they can be naturally extended for Riemann surfaces with the aid of local charts.

\subsection{Linear Beltrami solver (LBS)}
Lui et al.~\cite{lui2013texture} developed a linear method called the \emph{Linear Beltrami Solver} (LBS) for computing a quasi-conformal map $f(x,y) = u(x,y) + iv(x,y)$ with a given Beltrami coefficient $\mu(x,y) = \rho(x,y) + i \eta(x,y)$. The idea is to consider the real and imaginary parts in the Beltrami equation~\eqref{eqt:beltramieqt} separately:
\begin{equation}
 \rho(x,y) + i \eta(x,y) = \mu(x,y) = \frac{(u_x - v_y) + i(v_x + u_y)}{(u_x + v_y) + i(v_x - u_y)},
\end{equation}
from which we can express $v_x$ and $v_y$ as linear combinations of $u_x$ and $u_y$:
\begin{equation}\label{eqt:combination1}
\left\{ \begin{array}{cc}
  v_y &= \alpha_1 u_x + \alpha_2 u_y,\\
  -v_x &= \alpha_2 u_x + \alpha_3 u_y,\\
  \end{array} \right.
\end{equation}
where
\begin{equation}
\label{eqt:alpha123}
  \alpha_1 = \frac{(\rho -1)^2 + \eta^2}{1-\rho^2 - \eta^2}, \ \ 
  \alpha_2 = -\frac{2\eta}{1-\rho^2 - \eta^2}, \ \ 
  \alpha_3 = \frac{(\rho +1)^2 +\eta^2}{1-\rho^2 - \eta^2}.
\end{equation}
Similarly, we can express $u_x$ and $u_y$ as linear combinations of $v_x$ and $v_y$:
\begin{equation}\label{eqt:combination2}
\left\{  \begin{array}{cc}
  -u_y &= \alpha_1 v_x + \alpha_2 v_y,\\
  u_x &= \alpha_2 v_x + \alpha_3 v_y.\\
  \end{array}\right.
\end{equation}
Since $(v_y)_x + (-v_x)_y = 0$ and $(-u_y)_x + (u_x)_y = 0$, from Eq.~\eqref{eqt:combination1} and Eq.~\eqref{eqt:combination2} we have
\begin{equation}\label{eqt:BeltramiPDE}
\nabla \cdot \left(A \left(\begin{array}{c}
u_x\\
u_y \end{array}\right) \right) = 0\ \ \mathrm{and}\ \ \nabla \cdot \left(A \left(\begin{array}{c}
v_x\\
v_y \end{array}\right) \right) = 0,
\end{equation}
where $A = \left( \begin{array}{cc} \alpha_1 & \alpha_2\\
\alpha_2 & \alpha_3 \end{array}\right)$. In the discrete case, Eq.~\eqref{eqt:BeltramiPDE} can be discretized as two sparse symmetric positive definite linear systems. Therefore, one can easily obtain $u_x, u_y, v_x, v_y$ (and hence the quasi-conformal map $f$) for any given $\mu$ by solving two linear systems with certain boundary constraints (see~\cite{lui2013texture} for details). We denote the above procedure by $f = \textbf{LBS}(\mu)$.

\section{Proposed methods}\label{sect:main}
Below, we first develop an efficient algorithm for conformally parameterizing an open surface with one hole onto a planar annulus. We then utilize this algorithm to develop another efficient method for conformally parameterizing a multiply-connected open surface with $k$ holes onto a unit disk with $k$ circular holes.

\subsection{Annulus conformal map for open surfaces with one hole}
\begin{figure}[t]
    \centering
    \includegraphics[width=\textwidth]{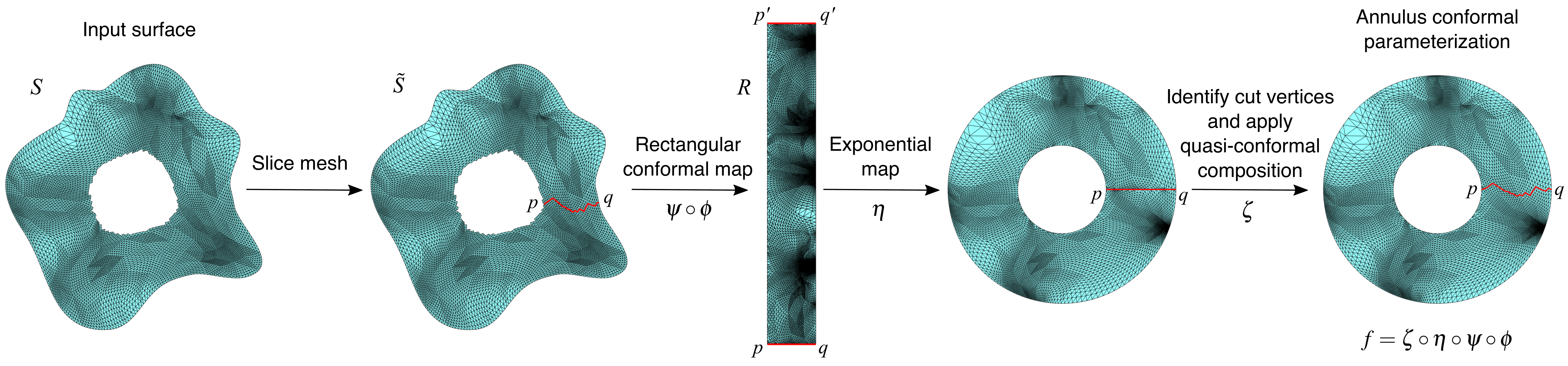}
    \caption{An illustration of the proposed annulus conformal map (ACM) method for open surfaces with one hole. We first slice the mesh along a path (highlighted in red) from the inner boundary to the outer boundary to make the surface open. We then map it onto a rectangle with an optimal length $L$ and unit width. The rectangle is subsequently mapped to an annulus using an exponential map. Finally, we identify the cut vertices and compose the map with another quasi-conformal map to achieve a conformal parameterization.}
    \label{fig:annulus_illustration}
\end{figure}

Let $S$ be an open surface in $\mathbb{R}^3$ with one hole, i.e. a topological annulus. Denote the surface boundary as $\partial S = \gamma_0 - \gamma_1$, where $\gamma_0$ is the outer boundary and $\gamma_1$ is the inner boundary. Our goal is to find a conformal parameterization $f:S \to \mathbb{C}$ that maps $S$ to an annulus on the plane with unit outer radius. The proposed method is outlined in Fig.~\ref{fig:annulus_illustration}.

To begin, we take an arbitrary vertex at the inner boundary $\gamma_1$ and find a shortest path from it to the outer boundary $\gamma_0$. By slicing $S$ along the path, we obtain a simply-connected open surface $\tilde{S}$ (see the red curve in Fig.~\ref{fig:annulus_illustration}, second left). Due to the change in the surface topology, it is possible to map $\tilde{S}$ onto a planar domain without holes. As for the reason of using a shortest path, note that the cut path between the two vertices corresponds to the two boundary edges connecting the left and right corners of the rectangle as illustrated in Fig.~\ref{fig:annulus_illustration}. Since the two edges are the geodesics between the corners, the shortest path between the two corresponding vertices in the input mesh is a natural choice for the cut path. Also, as the corners of the rectangle are enforced to be with angle $=\pi/2$, using a shortest path can help reduce the angular distortion there in the discrete case.

Now, we consider mapping $\tilde{S}$ onto a strip conformally (see Fig.~\ref{fig:annulus_illustration}, middle). Meng et~al.~\cite{meng2016tempo} developed an efficient rectangular conformal mapping algorithm based on the LBS method~\cite{lui2013texture}. The algorithm first computes an initial flattening map of the input surface onto the unit disk. It then maps the disk to the unit square using the LBS method. In particular, four boundary vertices are chosen as the four corners of a unit square, and an optimal quasi-conformal map is computed for mapping the remaining vertices onto the square domain. Finally, it keeps the length of the square domain fixed and optimally rescale the width of it so as to achieve a rectangular conformal map. Here, we follow the approach in~\cite{meng2016tempo} with some modifications for obtaining the conformal map onto a strip. 

We first compute a disk harmonic map $\phi:\tilde{S} \to \mathbb{D}$ by solving the following Laplace equation
\begin{equation}
\left\{
\begin{array}{c}
    \Delta \phi = 0,\\
    \phi(\partial \tilde{S}) = \partial \mathbb{D},
    \end{array} \right.
\end{equation}
where the boundary constraint is given by the arc-length parameterization. More explicitly, denote $\{p_i\}_{i=1}^n$ as the boundary vertices of $\tilde{S}$ in anti-clockwise order. For every $i$, we map $p_i$ to a point $(\cos\theta_i, \sin \theta_i)$ on the unit circle, where $\theta_1 = 0$, and $\theta_2, \theta_3, \dots, \theta_n$ are determined using the boundary edge lengths:
\begin{equation}
    \theta_i = \frac{2\pi \sum_{j=1}^{i-1} l_{[p_j, p_{j+1}]}}{\sum_{j=1}^{n} l_{[p_j, p_{j+1}]}}, \ \ i = 2, 3, \dots, n.
\end{equation}
Here, $l_{[p_j, p_{j+1}]}$ is the length of the edge $[p_j, p_{j+1}]$, and $p_{n+1} = p_0$. The Laplacian $\Delta$ is the Laplace--Beltrami operator on $\tilde{S}$, which can be easily discretized using the cotangent formulation~\cite{pinkall1993computing}. After flattening the sliced surface onto the unit disk, we compute a quasi-conformal map $\psi:\mathbb{D} \to R = [0,L] \times [0,1]$ from the unit disk to a rectangular domain with length $L$ and unit width, where $L$ is to be determined. In particular, we use the LBS method with the target Beltrami coefficient being $\mu_\psi = \mu_{\phi^{-1}}$:
\begin{equation} \label{eqt:lbs_rect}
    \psi = \textbf{LBS}(\mu_{\phi^{-1}}),
\end{equation}
where the four vertices on $\partial \tilde{S}$ that correspond to the endpoints of the cut path are set to be the four corners of the target rectangular domain. More explicitly, denote $p, p'$ as the two vertices on $\partial \tilde{S}$ that correspond to the endpoint at the inner boundary $\gamma_1$, and $q, q'$ as the two vertices on $\partial \tilde{S}$ that correspond to the endpoint at the outer boundary $\gamma_0$. The four corners of $R$ are set as follows (see Fig.~\ref{fig:annulus_illustration}):
\begin{equation}
    \psi(\phi(p)) = (0,0), \psi(\phi(q)) = (L,0), \ \ \psi(\phi(q')) = (L, 1), \ \ \psi(\phi(p')) = (0, 1).
\end{equation}
Note that by Eq.~\eqref{eqt:composition}, the conformal distortion of the quasi-conformal composition $\psi \circ \phi$ can be significantly reduced given an appropriate boundary constraint. In the original formulation~\cite{meng2016tempo}, the boundary vertices are allowed to freely slice along the sides of the rectangular domain to achieve conformality. However, in our case, the top and bottom sides of $R$ correspond to the cut path and are with equal number of corresponding vertices (see the two red curves in Fig.~\ref{fig:annulus_illustration}, middle). To enforce their positional consistency, we impose a periodic boundary constraint on the $x$-coordinates of the top and bottom boundary vertices. As for the choice of $L$, we start with an initial guess $L=1$ and compute the map $\psi$ using Eq.~\eqref{eqt:lbs_rect}. Then, we search for the optimal $L$ which minimizes the norm of the Beltrami coefficient of $\psi \circ \phi$ to further reduce the conformal distortion. 

Here we remark that one may look for an extra shear transformation $\begin{pmatrix} x \\ y \end{pmatrix} \mapsto \begin{pmatrix} 1 & 0 \\ a & 1 \end{pmatrix} \begin{pmatrix} x \\ y \end{pmatrix}$  to transform $R$ into a parallelogram, such that the two bottom corner points do not necessarily have the same $y$-coordinates. Theoretically, this can help further reduce the conformal distortion of the mapping. However, as the cut path is chosen to be a shortest path, we find that the optimal $a$ is usually very small (with $|a| \sim 10^{-4}$) in our experiments and the improvement in the conformality is negligible. Therefore, this step can be skipped in practice.

After getting the rectangular parameterization $\psi \circ \phi$, we apply the exponential map
\begin{equation}
    \eta(z) = e^{2\pi (z-L)},
\end{equation}
which maps the rectangular domain $[0,L] \times [0,1]$ to an annulus with inner radius $e^{-2\pi L}$ and outer radius 1. Because of the periodicity imposed in the computation of the rectangular parameterization, the top and bottom boundaries (i.e. the cut path vertices) are mapped to consistent locations on the annulus domain. We can then identify every pair of them and obtain a seamless mapping result (see Fig.~\ref{fig:annulus_illustration}, second right).

Finally, we use the quasi-conformal composition to further improve the conformality of the annulus map. Specifically, we compute an automorphism $\zeta$ on the annulus $(\eta \circ \psi \circ \phi)(S)$ with
\begin{equation} 
    \zeta = \textbf{LBS}(\mu_{(\eta \circ \psi \circ \phi)^{-1}}),
\end{equation}
where all boundary vertices are fixed. This results in the final annulus conformal parameterization $f = \zeta \circ \eta \circ \psi \circ \phi$ (see Fig.~\ref{fig:annulus_illustration}, rightmost). We remark that by the quasi-conformal composition, the Beltrami coefficient of the resulting map is with supremum norm less than 1 and hence is bijective.

The proposed annulus conformal map (ACM) algorithm is summarized in Algorithm~\ref{alg:annulus}.

\begin{algorithm}[h]
\KwIn{An open surface $S$ with annulus topology.}
\KwOut{A conformal parameterization $f:S \to \mathbb{C}$ onto an annulus with unit outer radius.}
\BlankLine
Compute a shortest path from an arbitrary vertex at the inner boundary to the outer boundary. Slice the mesh along the path\;
Compute the disk harmonic map $\phi$ for initialization\;
Compute the rectangular conformal map $\psi$ with a periodic boundary constraint, where the four corners correspond to the endpoints of the cut path\;
Apply the exponential map $\eta$ to obtain an annulus with unit outer radius\;
Compose the map with another quasi-conformal map $\zeta$ to further improve the conformality\;
The resulting conformal parameterization is given by $f = \zeta \circ \eta \circ \psi \circ \phi$\;
\caption{Annulus conformal map (ACM) for open surfaces with one hole.}
\label{alg:annulus}
\end{algorithm}

\subsection{Poly-annulus conformal parameterization of multiply-connected open surfaces with $k$ holes}
\begin{figure}[t]
    \centering
    \includegraphics[width=\textwidth]{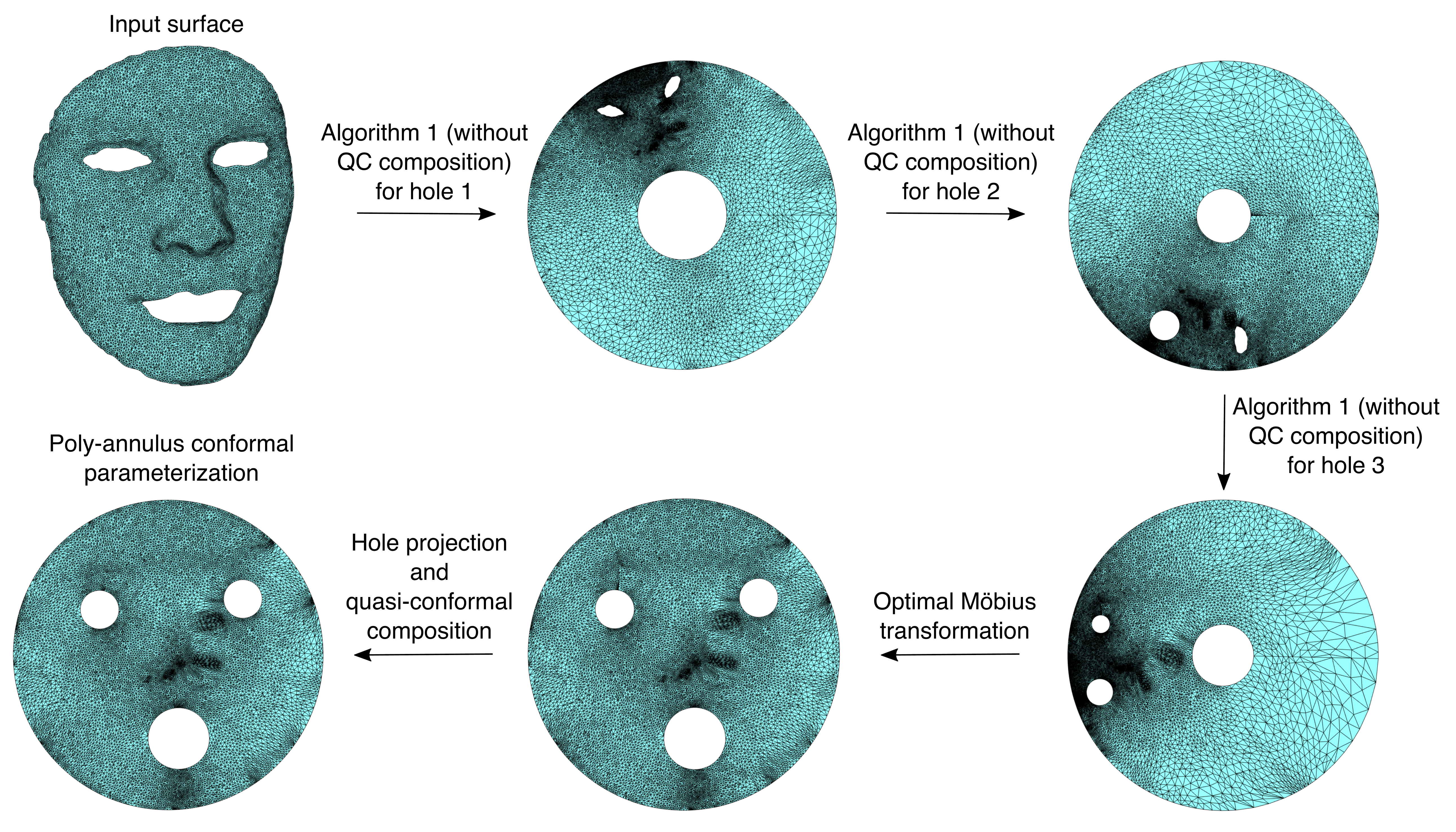}
    \caption{An illustration of the proposed poly-annulus conformal map (PACM) method for multiply-connected open surfaces. We first repeatedly apply Algorithm~\ref{alg:annulus} (without the quasi-conformal composition step) with all but one holes filled. After handling all holes, we compute an optimal M\"obius transformation to adjust the location of the holes. Note that the holes are all close to circles but may not be perfectly circular in practice. Finally, we further apply a projection step for enforcing the circularity of the holes and then compose the map with another quasi-conformal map to achieve a conformal parameterization.}
    \label{fig:polyannulus_illustration}
\end{figure}

Let $S$ be a multiply-connected open surface in $\mathbb{R}^3$ with $k$ holes, i.e. a topological poly-annulus. Denote the surface boundary as $\partial S = \gamma_0 - \gamma_1 - \gamma_2 - \cdots - \gamma_k$, where $\gamma_0$ is the outer boundary and $\gamma_1, \cdots, \gamma_k$ are the inner boundaries. Our goal is to find a conformal parameterization $f:S \to \mathbb{D}$ that maps $S$ to the unit disk with $k$ circular holes. The proposed method is outlined in Fig.~\ref{fig:polyannulus_illustration}.

Analogous to the Koebe's iteration method~\cite{koebe1910konforme,zeng2009generalized}, our method handles the $k$ holes of the surface $S$ one by one. We first fill all but the first holes to get a surface $S_1$ with annulus topology. In practice, one can simply fill a hole by adding a new vertex at the center of the hole and including its one-ring neighborhood of triangular faces. We can then apply the proposed ACM method (Algorithm~\ref{alg:annulus}) with the the quasi-conformal composition step (Line 5 in Algorithm~\ref{alg:annulus}) skipped to obtain an annulus map $g_1:S_1 \to \mathbb{C}$, with $\gamma_0$ and $\gamma_1$ mapped to the outer circle and the inner circle respectively. After that, we remove all filled regions to restore the surface topology. Here, we remark that the quasi-conformal composition step is skipped for simplifying the computational procedure. As shown in Fig.~\ref{fig:annulus_illustration}, the quasi-conformal composition primarily improves the conformality near the cut path, while the conformality of all other regions is largely unaffected. Therefore, we can leave the quasi-conformal composition step to the last part of the poly-annulus parameterization, which allows us to correct the conformal distortion in one solve instead of $k$ solves. 

By repeating the above process for handling all the remaining holes, we obtain the composition map $g_k \circ g_{k-1} \circ \cdots \circ g_1$. Here, note that each time one new hole is enforced to be a perfect circle, and in fact the circularity of the previously handled holes is also largely preserved. The reason is that the hole filling procedure involves adding a new vertex at the center of those holes together with a ring of triangles, and hence each of the subsequent annulus maps (which is highly conformal at most places even without the quasi-conformal composition step) will preserve the angles in the entirely filled shape including those in the filled holes as much as possible, thereby effectively preventing the previously handled holes from being largely distorted in circularity. 

\begin{figure}[t]
    \centering
    \includegraphics[width=0.8\textwidth]{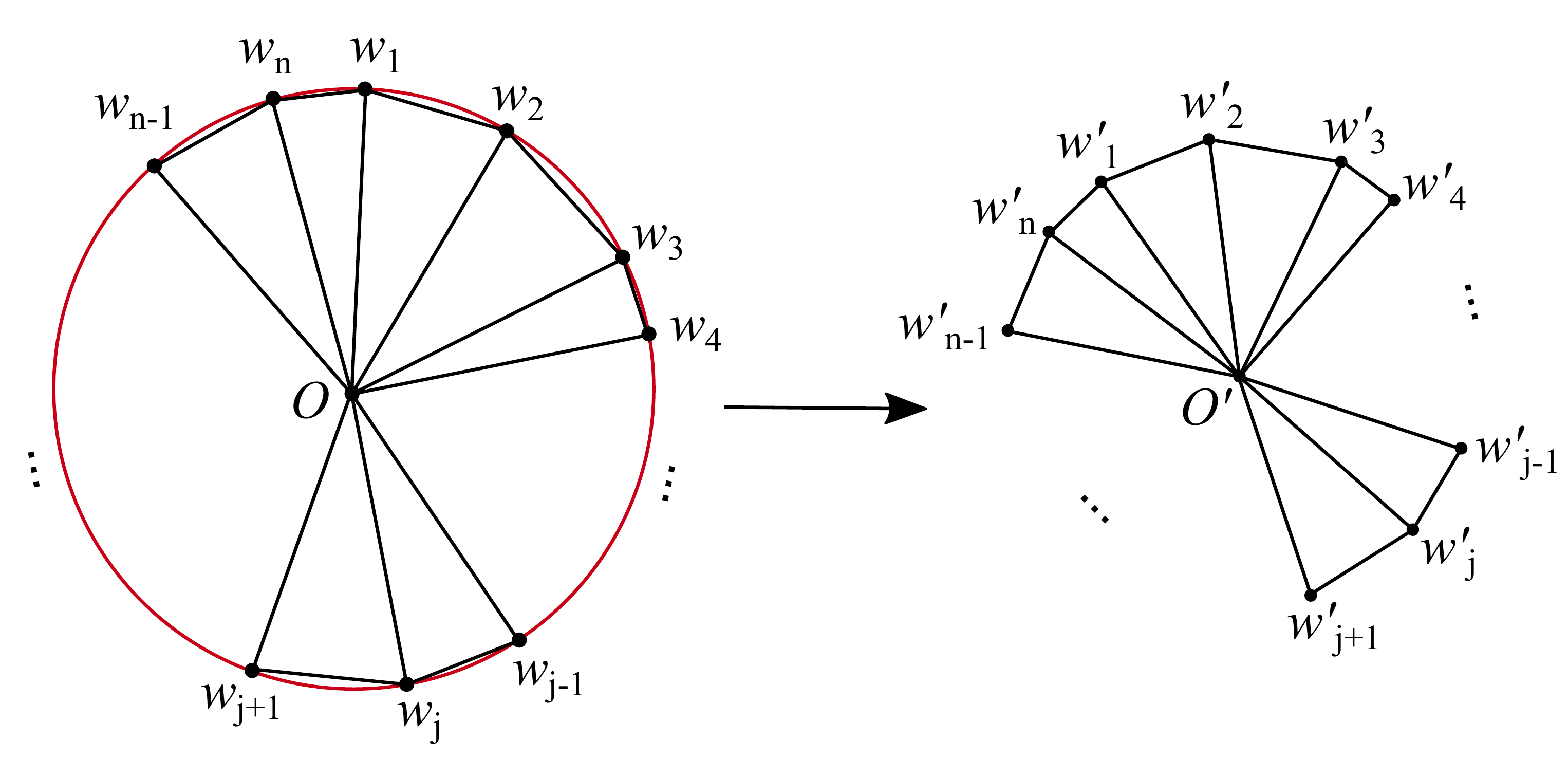}
    \caption{For a hole which has already been mapped to a circle at a previous step (left), the newly added vertex $O$ at the center of the hole at the next hole filling step is equidistant from all the boundary vertices $w_1, w_2, \dots, w_n$ and hence we have $\angle Ow_j w_{j+1} = \angle Ow_{j+1} w_{j}$ for all $j$. Under the next annulus map, the angles in the ring of triangles remain largely unchanged (right), i.e. $\angle O'w_j' w_{j+1}' \approx \angle Ow_j w_{j+1} = \angle Ow_{j+1} w_{j} \approx \angle O'w_{j+1}' w_{j}'$ for all $j$. Hence $O'$ is approximately equidistant from all $w'_j$, which indicates that the hole remains to be close to a circle under the mapping.}
    \label{fig:circularity}
\end{figure}

More specifically, consider a hole which has been mapped to a small circle and denote its vertices as $w_1, w_2, \dots, w_n$. Since the hole is circular, the new vertex added to that hole in the next hole filling procedure (denoted as $O$) is the center of the small circle (see Fig.~\ref{fig:circularity}, left), which implies that $O$ is equidistant from all $w_j$:
\begin{equation}
    l_{[O, w_1]} = l_{[O, w_2]} = \dots = l_{[O, w_n]}.
\end{equation}
It is then straightforward to see that $\angle Ow_j w_{j+1} = \angle Ow_{j+1} w_{j}$ for all $j$ in the newly added ring of triangles. Now, suppose the vertices are mapped to $w_1', w_2', \dots, w_n', O'$ under the next annulus map (see Fig.~\ref{fig:circularity}, right). As the annulus map is highly conformal, for every $j$ we should have 
\begin{equation}
    \angle O'w_j' w_{j+1}' \approx \angle Ow_j w_{j+1} = \angle Ow_{j+1} w_{j} \approx \angle O'w_{j+1}' w_{j}',
\end{equation}
and hence $l_{[O', w_j']} \approx l_{[O', w_{j+1}']}$ for all $j$ . It follows that $O'$ is approximately equidistant from all $w_j'$, which implies that the hole formed by $w_1', w_2', \dots, w_n'$ is close to a circle. One can then repeat the above argument at all subsequent steps to show that the hole will continue to be approximately circular. In other words, by the property of the hole filling procedure and the annulus map, every hole will remain approximately circular once it has been enforced to be a circle at a certain step. While the holes may not all be perfectly circular as numerical errors may accumulate throughout the entire process, an extra step of further enforcing the circularity will be added later.

Now, note that the $k$-th hole of $S$ is mapped to the center of the unit disk. Since this may not follow the distribution of the holes on $S$ well, the area distortion of the map may be large. To alleviate this issue, we use the M\"obius area correction scheme~\cite{choi2020parallelizable} to reduce the area distortion of the map while preserving conformality, thereby ensuring that the holes are at appropriate locations on the planar domain. More explicitly, we search for an optimal automorphism $\tau_{\alpha}$ on the unit disk in the following form:
\begin{equation} \label{eqt:mobius_disk}
\tau_{\alpha}(z) = \frac{z-\alpha}{1-\overline{\alpha} z},
\end{equation}
where $\alpha \in \mathbb{C}$ with $|\alpha|<1$, such that the composition $\tau_{\alpha} \circ g_k \circ g_{k-1} \circ \cdots \circ g_1$ minimizes the area distortion of the parameterization with respect to the input surface. 

As discussed above, all the $k$ holes become close to circles under the annulus mapping steps but they may not all be perfectly circular. Also, while M\"obius transformations map circles and straight lines to circles and straight lines in theory, in the discrete case they may cause a small distortion in the circularity of the holes. Therefore, we add a step of enforcing the circularity of the holes via projections. More specifically, for each hole $(\tau_{\alpha} \circ g_k \circ g_{k-1} \circ \cdots \circ g_1)(\gamma_i)$, we find the maximum inscribed circle of it and project all boundary vertices onto this circle. After performing this operation for all $k$ holes, we obtain a unit disk with $k$ circular holes. We denote the process by $\rho: \mathbb{D} \to \mathbb{D}$.

Finally, we use the quasi-conformal composition to further reduce the conformal distortion caused by the annulus mapping steps and the projection step. We compute an automorphism $h$ on the unit disk with the Beltrami coefficient $\mu_{(\rho \circ \tau_{\alpha} \circ g_k \circ g_{k-1} \circ \cdots \circ g_1)^{-1}}$ using the LBS method~\cite{lui2013texture}: 
\begin{equation}
    h = \textbf{LBS}(\mu_{(\rho \circ \tau_{\alpha} \circ g_k \circ g_{k-1} \circ \cdots \circ g_1)^{-1}}),
\end{equation}
where all boundary vertices are fixed. By the composition formula in Eq.~\eqref{eqt:composition}, the composition $f = h \circ \rho \circ \tau_{\alpha} \circ g_k \circ g_{k-1} \circ \cdots \circ g_1$ gives a conformal parameterization of $S$ onto the unit disk with exactly $k$ circular holes. Similar to Algorithm~\ref{alg:annulus}, the quasi-conformal composition here also ensures that the Beltrami coefficient of the resulting map is with supremum norm less than 1 and hence is bijective.

The proposed conformal parameterization method for poly-annulus surfaces is summarized in Algorithm~\ref{alg:poly_annulus}.

\begin{algorithm}[h]
\KwIn{A multiply-connected open surface $S$ with $k\geq 1$ holes.}
\KwOut{A conformal parameterization $f:S \to \mathbb{C}$ onto a unit disk with $k$ circular holes.}
\BlankLine
\For{$i = 1, \dots, k$ }{
Fill all but the $i$-th holes\;
Solve for an annulus map $g_i$ using Algorithm~\ref{alg:annulus} with the quasi-conformal composition step skipped\;
Remove all filled regions\;
}
Search for an optimal M\"obius transformation $\tau_{\alpha}$ for reducing the area distortion\;
Enforce the circularity of the holes by a projection step $\rho: \mathbb{D} \to \mathbb{D}$\;
Compose the map with another quasi-conformal map $h$ to improve the conformality\;
The resulting parameterization is given by $f = h \circ \rho \circ \tau_{\alpha} \circ g_k \circ g_{k-1} \circ \cdots \circ g_1$\;
\caption{Poly-annulus conformal map (PACM) for multiply-connected open surfaces.}
\label{alg:poly_annulus}
\end{algorithm}

\section{Experimental results} \label{sect:results}
The proposed conformal parameterization algorithms are implemented in MATLAB. The linear systems are solved using the backslash operator ($\backslash$) in MATLAB. For the step of mesh slicing in Algorithm~\ref{alg:annulus}, we use the MATLAB function \texttt{graphshortestpath} to compute a shortest path between the arbitrary vertex at the inner boundary $\gamma_1$ and the closest vertex at the outer boundary $\gamma_0$ of the input triangular mesh. For the rectangular conformal map in Algorithm~\ref{alg:annulus}, we use the MATLAB function \texttt{fminbnd} to search for the optimal length $L$ of the rectangular domain. For the M\"obius transformation in Eq.~\eqref{eqt:mobius_disk} in Algorithm~\ref{alg:poly_annulus}, we write $\alpha = re^{i\theta}$ and use the MATLAB function \texttt{fmincon} to search for the optimal parameters $(r,\theta) \in [0,1] \times [0, 2\pi]$. For the step of finding maximum inscribed circles of the holes, we use the function \texttt{find\_inner\_circle} available in the MATLAB Central FileExchange~\cite{findinnercircle}. All experiments are performed on a PC with a 4.0~GHz quad core CPU and 16~GB RAM.

To assess the conformality of a parameterization $f:S \to \mathbb{C}$, we consider the angular distortion of every angle $[v_i,v_j,v_k]$ on the surface mesh under the parameterization:
\begin{equation}\label{eqt:angle_distortion}
    d([v_i,v_j,v_k]) = \angle [f(v_i), f(v_j), f(v_k)] - \angle [v_i, v_j, v_k].
\end{equation}
For an ideal conformal map, we should have $d = 0$ for all angles. 

\begin{figure}[t!]
    \centering
    \includegraphics[width=\textwidth]{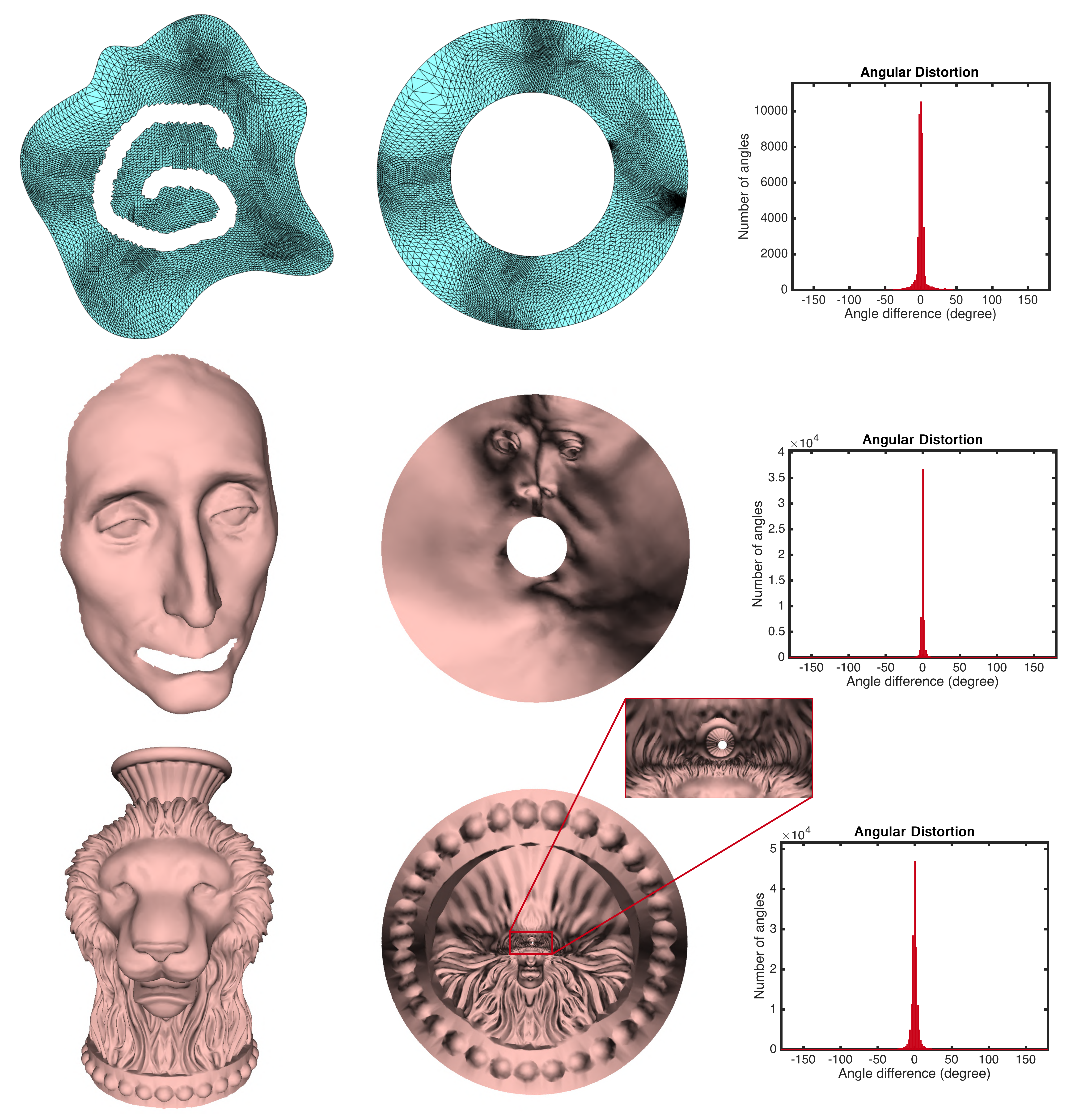}
    \caption{Examples of the annulus conformal parameterization achieved by our proposed ACM method (Algorithm~\ref{alg:annulus}). Left: The input open surfaces with annulus topology. Middle: The annulus conformal parameterization results. Right: The histograms of the angular distortion $d$.}
    \label{fig:result_acm}
\end{figure}

\begin{figure}[t!]
    \centering
    \includegraphics[width=\textwidth]{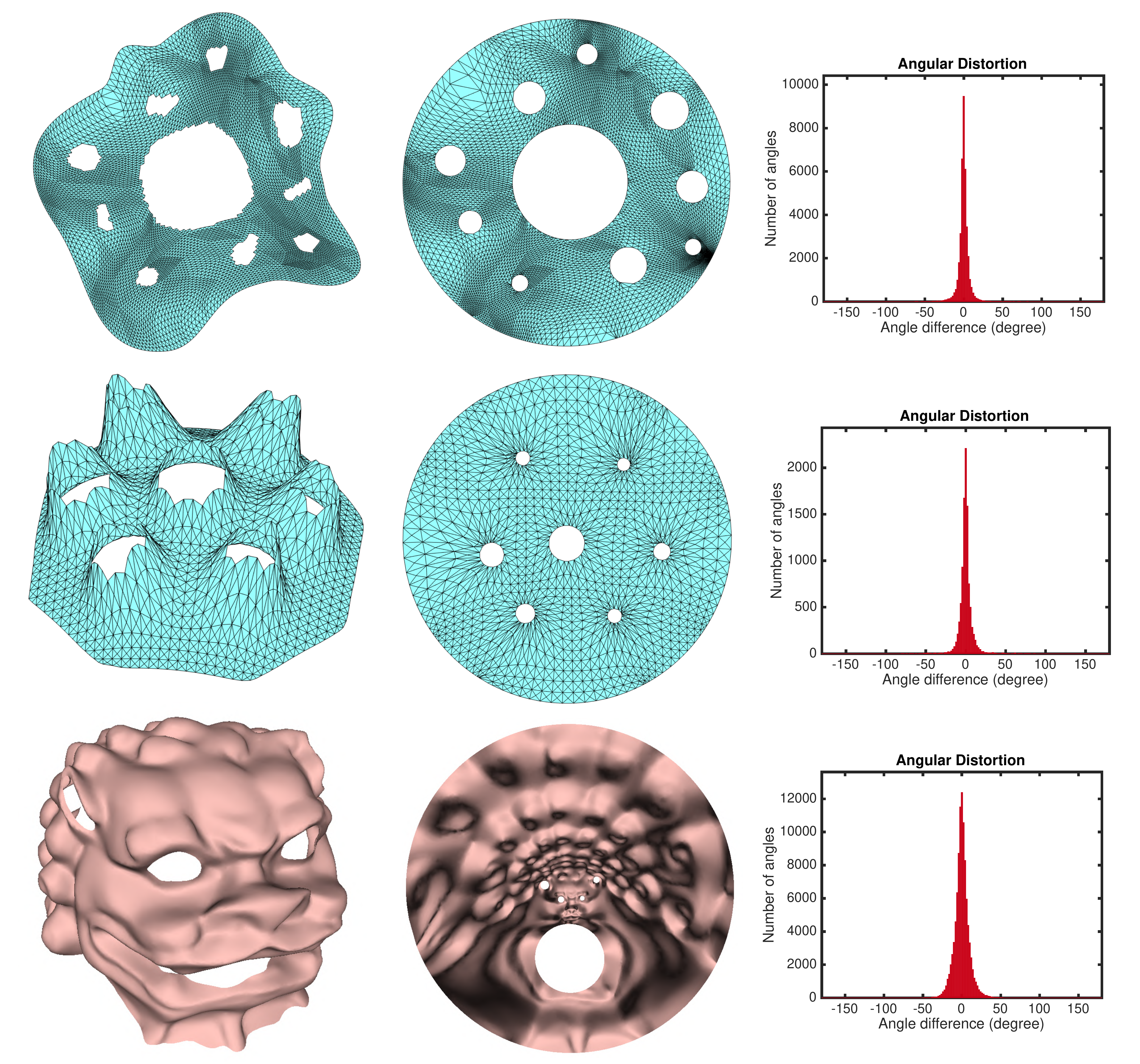}
    \caption{Examples of the poly-annulus conformal parameterization achieved by our proposed PACM method (Algorithm~\ref{alg:poly_annulus}). Left: The input multiply-connected open surfaces with $k$ holes. Middle: The poly-annulus conformal parameterization results. Right: The histograms of the angular distortion $d$.}
    \label{fig:result_pacm}
\end{figure}

Fig.~\ref{fig:result_acm} shows several surfaces with annulus topology and the annulus conformal parameterizations achieved by our proposed ACM method. Note that our method is capable of handling surfaces with a highly non-convex hole (see the top example) as well as surfaces with a highly tubular geometry (see the bottom example). From the histograms of the angular distortion $d$, it can be observed that the parameterizations are highly conformal. For comparison, we consider the Ricci flow (RF) method~\cite{jin2008discrete} with implementation available in the RiemannMapper toolbox~\cite{riemannmapper}. Table~\ref{table:annulus} records the computation time and the conformal distortion of our proposed ACM method and the RF method, from which it can be observed that our method outperforms the RF method in both the conformality and efficiency. The improvement in the conformality can be explained by that the RF method is based on the circle packing metric, which is highly dependent on the quality of the triangulations. For surface meshes with coarse or irregular triangulations, the approximation of the metric may be inaccurate, thereby yielding a large conformal distortion in the parameterization results. By contrast, our method effectively reduces the conformal distortion using the idea of quasi-conformal composition. As for the improvement in the efficiency, note that the RF method uses a gradient descent approach for minimizing the Ricci energy and hence is time-consuming, while our method only involves solving a few linear systems and a one-dimensional optimization problem.

\begin{table}[t!]
    \begin{center}
    \begin{tabular}{ |c|c|c|c|c|c| }
    \hline
    \multirow{ 2}{*}{Surface} & \multirow{ 2}{*}{\# vertices} & \multicolumn{2}{ c| }{ACM} & \multicolumn{2}{ c| }{RF~\cite{jin2008discrete}} \\
    \cline{3-6}
    & & Time (s) & Mean($|d|$) & Time (s) & Mean($|d|$) \\ \hline
    Amoeba1 (Fig.~\ref{fig:annulus_illustration}) & 7K & 0.3 & 1.1  & 4.7 & 21.5 \\ \hline
    Amoeba2 (Fig.~\ref{fig:result_acm}) & 7K & 0.3 & 4.0   & 4.5 & 21.6 \\ \hline
    Niccol\`o (Fig.~\ref{fig:result_acm}) & 10K & 0.3 & 1.5   & 5.9 & 18.6  \\ \hline
    Sophie (Fig.~\ref{fig:illustration}) & 21K & 1.0 & 0.6   & 15.3 & 9.5  \\ \hline
    Lion vase (Fig.~\ref{fig:result_acm}) & 25K & 1.1 & 3.2  & 17.0 & 26.3  \\ \hline
    \end{tabular}
    \end{center}
    \caption{Performance of the proposed ACM method (Algorithm~\ref{alg:annulus}) and the Ricci flow (RF) method~\cite{jin2008discrete} for the annulus conformal parameterization of open surfaces with annulus topology. Here, the angular distortion $d$ is evaluated using Eq.~\eqref{eqt:angle_distortion}.}
    \label{table:annulus}
\end{table}

Fig.~\ref{fig:result_pacm} shows the poly-annulus conformal parameterizations of several multiply-connected open surfaces achieved by our proposed PACM method. It can be observed that our method works well for surfaces with different size, shape, and number of holes. For a more quantitative analysis, we again compare our method with the RF method in terms of the computation time and the angular distortion. As shown in Table~\ref{table:poly-annulus}, our method achieves a significant improvement in both the efficiency and conformality when compared to the RF method. Therefore, our method is more advantageous for the computation of the poly-annulus conformal parameterization.

\begin{table}[t!]
    \begin{center}
    \begin{tabular}{ |C{35mm}|c|c|c|c|c|c| }
    \hline
    \multirow{ 2}{*}{Surface} & \multirow{ 2}{*}{\# holes}  & \multirow{ 2}{*}{\# vertices} &  \multicolumn{2}{ c| }{PACM} &  \multicolumn{2}{ c| }{RF~\cite{jin2008discrete}} \\
    \cline{4-7}
    & &  &  Time (s) & Mean($|d|$) & Time (s) & Mean($|d|$)\\ \hline
    David (Fig.~\ref{fig:texture_mapping}) & 2 & 25K & 2.6 & 0.8    & 16.6 & 13.8  \\ \hline
    Alex (Fig.~\ref{fig:illustration}) & 3 & 14K & 2.0 & 1.3   & 10.5 & 13.1   \\ \hline
    Face (Fig.~\ref{fig:remeshing}) & 3 & 1K  & 0.2 & 4.4    & 0.8 & 14.1   \\ \hline
    Lion (Fig.~\ref{fig:result_pacm}) & 5 & 17K & 3.5 & 6.8 & 12.4 & 10.2 \\ \hline 
    Peaks (Fig.~\ref{fig:result_pacm}) & 7 & 2K & 0.4 & 5.1 & 1.4 & 15.8  \\ \hline
    Twisted hemisphere (Fig.~\ref{fig:texture_mapping}) & 8 & 25K & 8.8 & 7.6  & 22.0 & 9.1  \\\hline
    Amoeba (Fig.~\ref{fig:result_pacm}) & 10 & 7K & 2.0 & 4.3   & 5.8 & 21.7   \\ \hline
    \end{tabular}
    \end{center}
    \caption{Performance of the proposed PACM method (Algorithm~\ref{alg:poly_annulus}) and the Ricci flow (RF) method~\cite{jin2008discrete} for the poly-annulus conformal parameterization of multiply-connected open surfaces. Here, the angular distortion $d$ is evaluated using Eq.~\eqref{eqt:angle_distortion}.}
    \label{table:poly-annulus}
\end{table}

\section{Applications}\label{sect:application}
\subsection{Texture mapping}
\begin{figure}[t!]
    \centering
    \includegraphics[width=\textwidth]{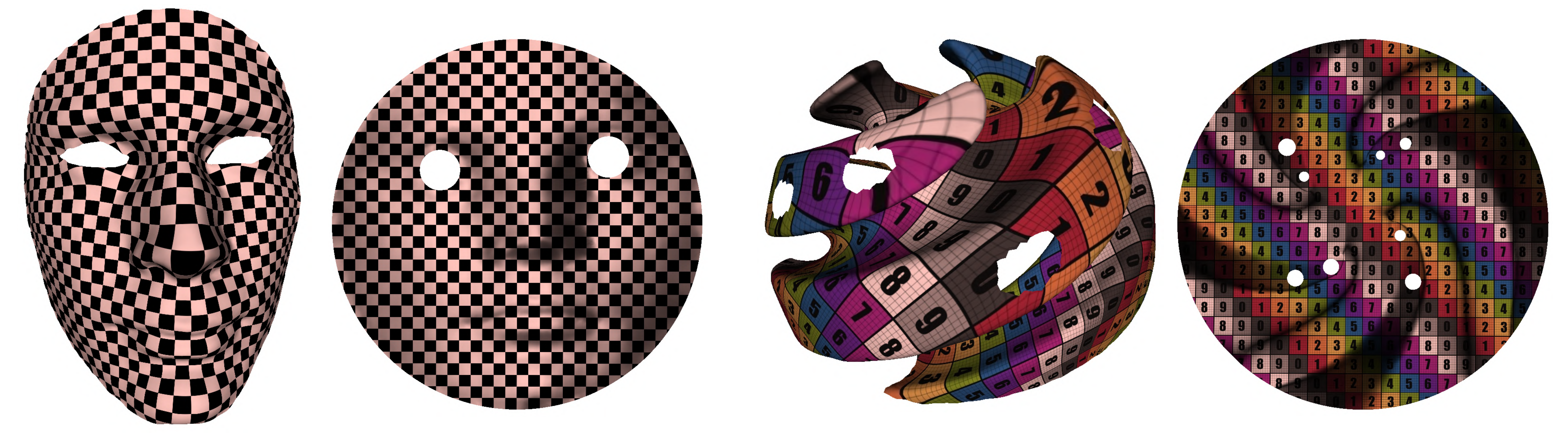}
    \caption{Texture mapping for multiply-connected surfaces achieved by our proposed conformal parameterization methods. We first compute the conformal parameterization of an input multiply-connected surface onto a planar domain using our proposed methods. Then, we can design textures on the planar domain and map them back onto the input surface with the local geometry well-preserved.}
    \label{fig:texture_mapping}
\end{figure}

The proposed conformal parameterization methods can be effectively applied to texture mapping. Using our methods, any multiply-connected open surface in $\mathbb{R}^3$ can be conformally mapped onto a unit disk with circular holes. Textures can then be designed on the plane and mapped back onto the surface easily. As our methods are angle-preserving, the local geometry of the designed textures will be well-preserved. Also, as our methods produce global parameterizations of the surfaces, the texture mapping results will be seamless. Fig.~\ref{fig:texture_mapping} shows two texture mapping results produced using our parameterization methods. The orthogonality of the checkerboard patterns on the surfaces indicates that our parameterizations are highly conformal.

\subsection{Surface remeshing}
The proposed conformal parameterization methods can also be applied to surface remeshing. Suppose we would like to improve the mesh quality of a given multiply-connected surface. A simple way is to map it onto the unit disk with circular holes using our parameterization methods, and then perform the remeshing process on the plane. 

\begin{figure}[t]
    \centering
    \includegraphics[width=\textwidth]{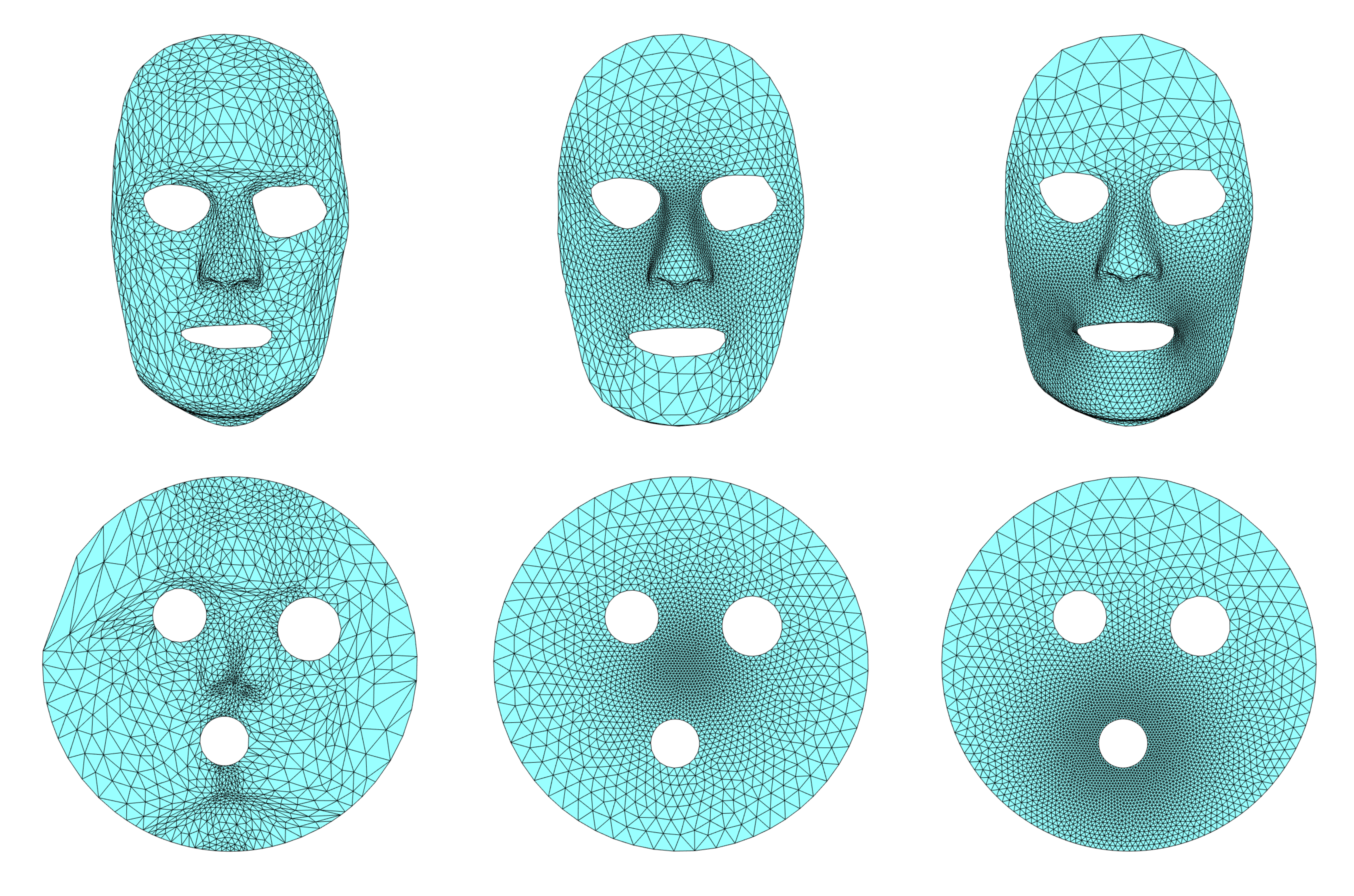}
    \caption{Remeshing a multiply-connected open surface via our proposed conformal parameterization methods. Given a multiply-connected open surface (top left), we first compute the poly-annulus conformal parameterization using our proposed methods (bottom left). We can then generate triangular meshes on the poly-annulus domain using DistMesh~\cite{persson2004simple} with different desired effects, such as having finer triangulations at the central part (bottom middle) or around one of the holes (bottom right). The new planar meshes can then be mapped back onto the given surface via the parameterization (top middle and top right).}
    \label{fig:remeshing}
\end{figure}

In particular, DistMesh~\cite{persson2004simple} is a powerful toolbox for generating triangular meshes, which uses signed distance functions to specify the geometry of the domain and control the mesh quality. In our case, the disk domain with circular holes can be easily expressed as a difference between signed distance functions for several circles using the \texttt{ddiff} and \texttt{dcircle} functions in DistMesh. The target mesh quality is set using a scaled edge length function in DistMesh, which can also be easily controlled using the signed distance functions for circles. Therefore, our parameterization methods can be naturally combined with DistMesh for the remeshing task. Once a new planar mesh is generated, we can map it back to the surface via the parameterization. Fig.~\ref{fig:remeshing} shows two examples of remeshing a multiply-connected human face surface, from which it can be observed that different remeshing effects can be easily achieved. As the parameterization is angle-preserving, the regularity of the triangles in the new planar meshes is well-preserved in the final remeshed surfaces. 

\begin{figure}[t]
    \centering
    \includegraphics[width=\textwidth]{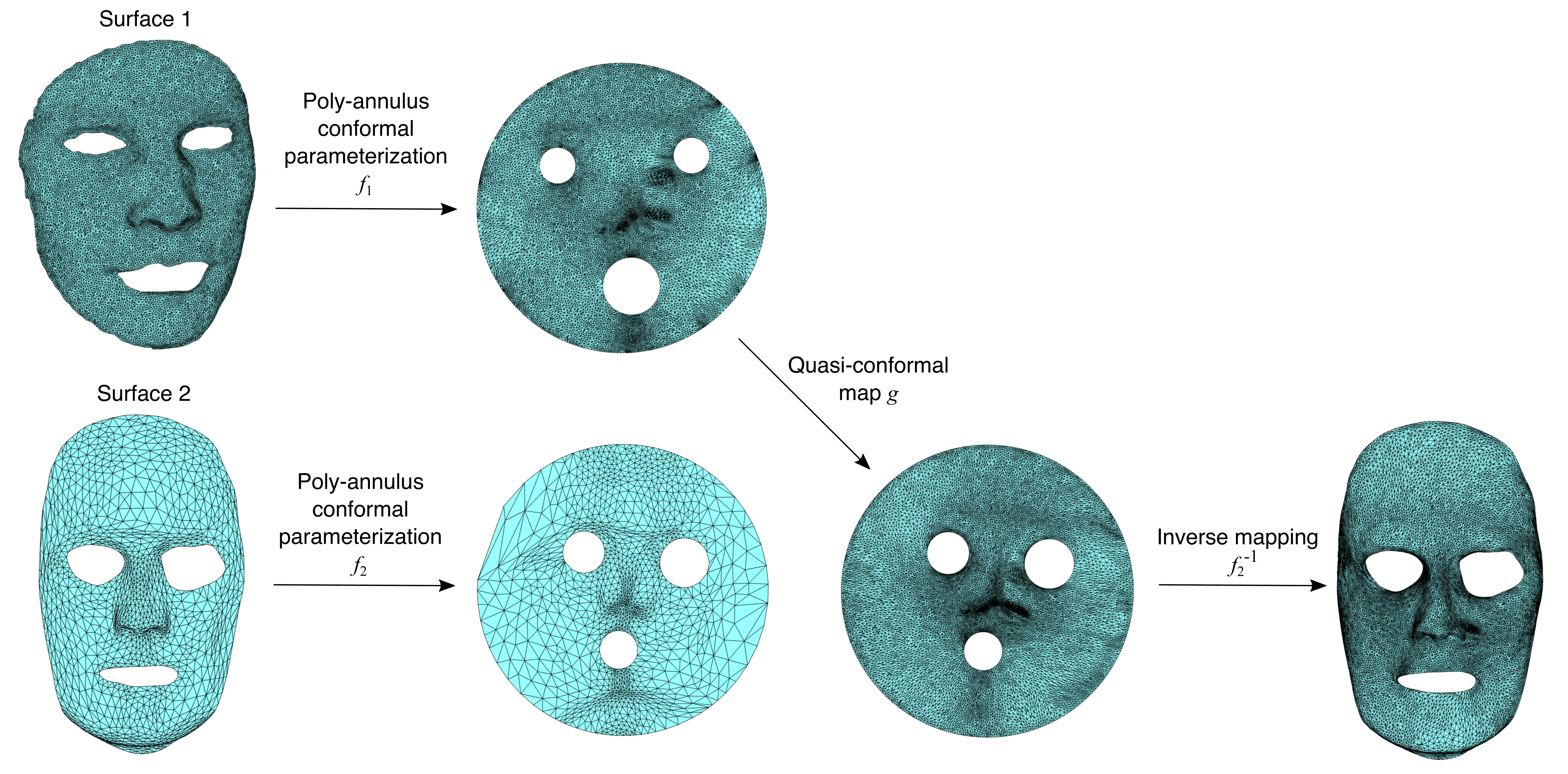}
    \caption{Registration of multiply-connected surfaces via the proposed conformal parameterization methods. Given two multiply-connected surfaces with the same topology, we can first conformally parameterize them onto two unit disk domains with the same number of circular holes. We can then compute a quasi-conformal map between the two planar domains, with all corresponding holes exactly matched. Finally, we can map the planar mapping result back onto the target surface to obtain the final registration result.}
    \label{fig:registration}
\end{figure}

\subsection{Surface registration}
Another possible application of the proposed conformal parameterization methods is multiply-connected surface registration~\cite{ng2014teichmuller}. As shown in Fig.~\ref{fig:registration}, given two multiply-connected open surfaces (denoted as $S_1$ and $S_2$) with the same topology, we can first compute the poly-annulus conformal parameterizations (denoted as $f_1$ and $f_2$) of them using our proposed methods. Then, we can compute a quasi-conformal map $g$ between the two planar domains $f_1(S_1)$ and $f_2(S_2)$ with all corresponding holes exactly matched using the LBS method. The composition $f_2^{-1} \circ g \circ f_1$ then gives a registration mapping between the surfaces $S_1$ and $S_2$. From the final registration result in Fig.~\ref{fig:registration}, it can be observed that the two multiply-connected surfaces are matched very well.

\section{Discussion} \label{sect:discussion}
With the advancement in computer technology, there has been a surge of interest in the development of conformal parameterization algorithms for science and engineering applications in recent decades. However, most of the existing methods only work for simply-connected surfaces. In this work, we have proposed two novel algorithms for the conformal parameterization of multiply-connected surfaces onto either an annulus or a unit disk with circular holes using quasi-conformal theory. As there are a vast number of analytical and numerical conformal mapping methods for multiply-connected planar domains~\cite{crowdy2005schwarz,crowdy2006conformal,crowdy2007schwarz,nasser2009numerical,nasser2020plgcirmap}, the proposed parameterization algorithms pave the way for applying these methods to multiply-connected surfaces. For instance, the prime function has been used for solving various applied and natural science problems on multiply-connected planar domains~\cite{crowdy2020solving}. With the aid of our proposed parameterization algorithms, it may be possible to extend the method to Riemann surfaces.

Besides the applications discussed in this work, it is natural to explore the use of the proposed conformal parameterization methods for shape analysis~\cite{zeng2008shape,zhao2019automatic}, greedy routing in sensor networks~\cite{li2015compact} etc. Also, note that both the proposed poly-annulus conformal parameterization method in this work and the conventional Koebe's iteration method rely on a series of annulus mappings for producing the final result. Alternatively, as shown in the recent discrete conformal equivalence approach~\cite{bobenko2016discrete}, the poly-annulus parameterization can be achieved by gluing faces to all but one boundary component and constructing a discrete conformal map using discrete conformal equivalence of cyclic polyhedral surfaces. A possible future direction would be to adopt a similar strategy to further simplify the computational procedure of our method. Another possible future direction is to combine the proposed methods with the optimal mass transport (OMT)~\cite{zhao2013area,su2016area,pumarola20193dpeople,giri2020open} or the density-equaling map (DEM)~\cite{choi2018density,choi2020area} for efficiently computing area-preserving parameterizations of multiply-connected surfaces.\\

\bibliographystyle{ieeetr}
\bibliography{multiplybib.bib}

\end{document}